# Generative Horcrux: Designing AI Carriers for Afterlife Selves

Lai, Zhen-Chi*[a,b]; Cheng, Yu-Ting[a]; Lin, Pei-Ying[c]; Ho, Chiao-Wei[d]; Huang, Janet Yi-Ching [b]

[a] Design Department, National Taiwan University of Science and Technology, Taipei, Taiwan
[b] Design Department, Eindhoven University of Technology, Eindhoven, the Netherlands
[c] HTI group in IE&IS, Eindhoven University of Technology, Eindhoven, the Netherlands
[d] Digital Medicine Lab, Taipei, Taiwan

* z.-.c.lai@tue.nl

As generative AI technologies rapidly advance, AI agents are gaining the ability not only to collect data and perform tasks but also to respond to environments and evolve over time. This shift opens new possibilities for reimagining digital legacy—raising critical questions about how we remember, commemorate, and interact with the traces of the deceased. The forms of these AI agents are particularly important, as they act as vessels for digital legacies—much like urns for ashes. We will ask: What kinds of devices or representations would we want to store our digital selves or legacies in? How do we envision future generations interacting with these forms? The question is not only about the function of these agents or the object's role as a storage vessel but also the meaning it carries, the memories it preserves, and its connection to the extended notion of our "Generative Horcrux." This three-hour in-person workshop invites design practitioners and researchers from diverse backgrounds to explore the emerging landscape of generative AI agent-based digital legacy. This workshop uses fiction and hands-on prototyping to explore how AI agents might reconfigure memory, identity, and posthumous presence in future sociotechnical worlds. We anticipate that this session will foster interdisciplinary dialogue and contribute conceptually and methodologically to HCI, design research, and AI ethics.

## 1    Workshop description

The rapid development of generative AI has given rise to AI agents as a novel medium for digital legacy (Lei et al., 2025), transforming them from passive repositories of personal data into interactive, posthumous actors. These agents are not merely containers of memory; they are capable of simulating personality traits, engaging in real-time interaction, and even evolving over time (Park et al., 2023). This marks a critical departure from traditional forms of digital legacy, such as digital photos, chat histories, and personal social media accounts, which typically preserve content in a static and non-interactive manner. Echoing the notion of "generative ghosts" proposed by Morris and Brubaker (2025), such generative AI agent-enhanced entities introduce a new form of posthumous presence that is agentic and adaptive, and even capable of participating in social and economic life. As such, they hold the potential to reshape how people grieve, how memories are preserved, and how the deceased are represented in society. However, while interest in such systems grows, there remains a critical gap in how they are materially and socially situated. We argue that the mediums and interfaces through which these agents are experienced, especially when they take physically embodied forms that integrate commemorative functions, play a critical role in shaping how memory is constructed, how users relate to digital content, and how these agents are socially positioned. These physical forms also have the potential to foster new kinds of attachment between the living and the digital dead (Orth et al., 2020), and to influence how individuals determine what constitutes meaningful data for preservation.

Imaginations of AI agent-based digital legacy have long appeared in science fiction—from memory playback devices (Barrett & Taylor, 2025), to voice assistants (Spike, 2013), and humanoid robotic replicas (Owen, 2013). Today, commercial services are beginning to bring these visions into reality: AI systems now co-create virtual memoirs with users (Kindered Tales, 2023; Life Story AI, 2023), preserve the voices of loved ones through voice simulation (HereAfter, 2023), and use image and language models to recreate historical figures (Kikumoto, 2023). Despite this momentum, most current implementations remain screen or voice-based, with limited exploration into how these AI agents might be meaningfully embodied. As AI begins to play a role in memory preservation and posthumous interaction, it becomes crucial to consider what kinds of physical forms or carriers these digital legacies might take. The material presence of AI agents—as objects, artifacts, or embodied systems—could significantly influence how they are perceived, remembered, and integrated into rituals of grief or memorialization. Previous HCI research on digital legacy has emphasized the importance of physical objects as triggers for memory and commemoration (Ankenbauer & Brewer, 2024; Beuthel & Fuchsberger, 2022; Hallam & Hockey, 2020; Odom et al., 2012). These works suggest that **material forms matter**: how digital memories are embedded in physical carriers—objects, devices, or embodied interfaces—can profoundly shape emotional attachment, memory construction, and interpersonal meaning-making. When generative AI Agents are introduced, the nature of interaction shifts fundamentally. These are no longer static memory objects for memorial attachment, but interactive entities capable of responding to the environment, generating new content, and asserting a kind of agency. This transformation raises a set of critical questions: As these memory bearing objects gain more agency, they may evolve from passive storage containers into interactive digital selves—entities with the ability to act, respond, and influence remembrance. This shift invites further exploration of how design can engage with the ethics, representation, and rituals of remembering through AI. To address these issues, we propose two research questions: (1) How might physical interfaces and embodied forms be designed to meaningfully support generative AI agent-based digital legacies and posthumous interaction with AI agents? (2) What ethical, emotional, and relational concerns arise when AI becomes a vessel for memory and remembrance?

Yet, the idea of an AI-embedded digital self still feels distant or abstract. To make this concept more tangible and personally relevant, we conduct design fiction strategies (Bleecker, 2022) to create a fictional design scenario in which they imagine themselves as already dead, and reflect on what kinds of carriers they want to have, and what traces, memories, and messages they would want to store inside. By envisioning their own AI afterlife, participants are invited into a speculative design experiment in which they prototype a carrier for their generative AI agent–based digital legacy—or, put differently, their Generative Horcrux. The term Horcrux, adapted from the Harry Potter series, refers to an object that holds a fragment of a wizard's soul, here reimagined as a metaphor for one's posthumous digital self. To ground this scenario in-personal relevance, participants are asked to bring a memory object that holds symbolic or emotional significance. This object serves as a prompt and design anchor, guiding them to write their own epitaph and to imagine a generative AI agent that would carry forward their presence in a speculative afterlife.This workshop invites designers and design researchers from diverse disciplines to collectively explore the future potential of AI agents as media for digital legacy. Participants will examine potential functions and forms of such agents, speculate on their emotional and societal implications, and critically reflect on the ethical challenges they present. We are particularly interested in how design can support interactions that are meaningful, culturally situated, and attentive to values, memory, and relational dynamics. We anticipate the following outcomes and contributions: (1) to generate interest and participation in this emerging topic and foster a growing research community around it, (2) to produce valuable participatory insights and speculative concepts that inspire new design research directions, and (3) to promote interdisciplinary dialogue about the societal and ethical implications of designing AI agents for life after death.

## 2 Workshop Structure

This workshop supports exploration of how generative AI agents may shape future digital legacies, particularly through critical engagement with the intersection of AI technologies and human mortality. The session will take place over three hours in-person and will include small group discussions, expert facilitation, and interdisciplinary exchange. By bringing together professionals in interaction design from diverse fields, the workshop seeks to spark dialogues and critical reflections on this emerging topic from multiple perspectives. To support active engagement, we plan to provide a series of design fiction materials that serve as prompts to stimulate imagination and participation. Due to time and material constraints, attendance will be limited to fewer than 20 participants to ensure deeper dialogue and high quality engagement. To facilitate a meaningful and structured workshop experience, we designed a fictional narrative scenario that invites participants to engage personally and empathetically with the speculative premise: imagining and designing their own digital afterlife. Through this narrative lens, the session is structured into five interconnected stages: **(1) Welcome to Life Legacy Design Agency—Enter the Company, Begin the Role, (2) Form Your Legacy Circle—Find the Others Who Remember You, (3) Build the Carrier—Design the Carrier of Your Afterlife (4) Return to the World—Present Your Posthumous Self, (5) Face the Ethics Committee—Defend Your Legacy.** Besides, in preparation for the workshop, participants will receive a short set of curated references to help them familiarize themselves with key concepts. These preparatory steps are intended to deepen participants' engagement and enrich the collective exploration during the workshop.

### 2.1 Pre-workshop Plan

Prior to the workshop, participants will be asked to review a set of assigned readings to develop a preliminary understanding of the topic. In addition, we will also invite them to bring a personal "memory object" that carries emotional or narrative significance. This object will serve as a starting point for imagining how generative AI agents might mediate memory and legacy.

## 2.2 Session Set up

The workshop is framed by a fictional narrative that invites participants to personally engage with the speculative task of designing their own AI agent-based digital legacy. It unfolds across five interconnected stages:

| | | |
|---|---|---|
| ~15 min | **Welcome to Life Legacy Design Agency.** *(Enter the Company, Begin the Role)* | |
| | "Welcome to Life Legacy Design Agency, a future-facing design firm specialising in AI afterlives." | |
| | At this stage, participants enter a role-play scenario: the workshop begins as an induction for new designers joining the agency. The organisers will introduce the firm's mission, products, and technologies related to digital legacy creation. Every participant is now a designer — responsible for developing new AI-based legacy services, either for clients or for themselves. Through this fictional frame, participants begin to imagine what it means to design digital afterlife. | |
| ~10 min | **Form Your Legacy Circle** *(Find the Others Who Remember You)* | |
| | "In this agency, every designer works as part of a team. Who will you collaborate with—and what stories will guide your project?" | |
| | Participants will be grouped based on shared meanings and themes found in their memory objects. These emerging "legacy circles" become the foundation for designing collective AI afterlife vessels. Within each group, participants reflect on their personal stories and begin to reimagine how traces of their identities might persist—together. | |
| ~70 min | **Build the Carrier** *(Design the Carrier of Your Afterlife)* | |
| | "Now, design your return. What will carry your voice across time?" | |
| | Participants will create prototypes that serve as vessels for their AI afterlives. These vessels may be conversational, embodied, ambient, or abstract—but they all must answer the same question: how can design mediate remembrance? Participants will use any materials or design tools they prefer, while organisers circulate to support ideation and framing. | |
| ~30 min | **Reveal the Legacy** *(Present Your Posthumous Legacy)* | |
| | "Your design is complete. Step forward and introduce the vessel you've created." | |
| | Each group selects one or two representatives to unveil their designed afterlife vessel to the full gathering. This session showcases how each prototype contributes to reimagining memory, identity, and legacy in the age of generative AI. Participants share the intentions, metaphors, and imagined use cases behind their ghost's design. | |
| ~55 min | **Face the Ethics Committee** *(Defend Your Legacy)* | |
| | "Before you linger in this world forever, your presence must be questioned." | |
| | This closing session convenes a speculative "ethics committee," composed of experts and participants, to reflect on the social, cultural, and moral implications of each design. Whose stories get told? Who consents? What are the risks of being remembered too well—or not at all? Through feedback, debate, and open discussion, the group collectively explores the responsibilities of designing for AI afterlife. | |

*Figure 1. A draft schedule of the day's activities*

## 2.3 Post-workshop Plan

The workshop will conclude with a collective reflection and summary discussion. With participants consent, we plan to document and potentially develop their designed artifacts into research-oriented design fiction materials to support future discussions in related areas of design research. Additional follow-up activities include the development of a workshop paper and an open source toolkit, allowing broader audiences to explore and reflect on their own potential digital legacies in the future.

## 3 Expected Outcome

This three-hour in-person workshop, to be held at IASDR 2025, invites designers and researchers from diverse backgrounds to collaboratively explore future imaginaries and design visions of generative AI–driven digital legacy. Grounded in participants' reflections and creative imagination, the workshop offers space for both conceptual inquiry and critical dialogue on the implications of AI-based legacy systems— particularly in relation to data privacy, authorship, and broader ethical concerns. In parallel, participants will engage in hands-on activities using emerging design research methods to prototype Generative Horcrux, speculative afterlife carriers. The workshop aims to generate a collection of creative responses, speculative prototypes, and conceptual insights that will inform ongoing research in this emerging field. These contributions will serve as reference points for future speculative design explorations around AI and digital legacy. We also aim to establish a reciprocal loop by sharing the design outcomes— initially inspired by participants— back with them after the event, enabling mutual inspiration and continued dialogue beyond the workshop. Outcomes from this workshop will feed into subsequent academic publications and creative works, contributing meaningfully to a growing conversation about posthumous digital identity and AI-mediated remembrance.

**About the Authors:**

**Lai, Zhen-Chi** is a double Ph.D. candidate at Eindhoven University of Technology and National Taiwan University of Science and Technology researching generative AI, non-human agency (Lai, Huang, & Liang, 2024; Lai, Tsai, & Liang, 2024), and human–AI collaboration. Through storytelling strategy, she explores how generative AI systems influence daily life and enable new forms of social futures.

**Cheng, Yu-Ting** is an Assistant Professor in Design Department at National Taiwan University of Science and Technology. Her research explores human–nonhuman–object relations through a thing-centered perspective (Cheng et al., 2023), and she develops co-design fiction workshops to engage the public with future-oriented topics (Cheng et al., 2018).

**Lin, Pei-Ying** (http://peiyinglin.net) is an artist, designer, and designer researcher in Science/Art who works with More-than-Human others including viruses, bacteria, textile (Lin, Andersen, Schmidt, Schoenmakers, Hofmeyer, Pauwels, & IJsselsteijn, 2024), and AI (Lin, Andersen, Schmidt, Schoenmakers, Hofmeyer, Pauwels, & Ijsselsteijn, 2024). She is currently a PhD at Eindhoven University of Technology, resident artist of RepliFate, and IF Fellow of Royal Shakespeare Company.

**Ho, Chiao-Wei** is a designer, publisher, educator, and founder of Digital Medicine Lab—an independent research lab focused on speculative design, live coding, and critical technologies. His work has been instrumental in advancing speculative design in Taiwan and establishing it as a critical approach within contemporary design practice.

**Huang, Janet Yi-Ching** is an Assistant Professor in Industrial Design at Eindhoven University of Technology. Her work bridges HCI and AI in design, creating methods and toolkits that help designers engage with data, algorithms, and hybrid intelligence as design materials (Huang et al., 2021).